\documentclass[aps,prl,superscriptaddress,reprint]{revtex4-1}

\usepackage{amsmath}
\usepackage{amssymb}
\usepackage{graphicx}
\usepackage{amsfonts}
\usepackage{bm}
\usepackage{mathrsfs}
\usepackage[mathscr]{eucal}
\usepackage{esint}
\usepackage{subfigure}
\usepackage{lineno}
\usepackage{esint}
\usepackage{subfigure}
\usepackage{supertabular}
\usepackage{booktabs}
\usepackage{amssymb}
\usepackage{latexsym}
\usepackage{CJK}
\usepackage{times}


\begin{document}
\hyphenation{along dif-ferent equation equ-ation model nature theory space speaking stand}

\title{Quantum Severalty: Speed-up and Suppress Effect in Searching Problem}
\author{Jin-Hui Zhu}
\affiliation{\noindent
Zhejiang Province Key Laboratory of Quantum Technology $\&$ Device and Department of Physics, Zhejiang University, Hangzhou 310027, P.R. China.}
\author{Li-Hua Lu}
\affiliation{\noindent
Zhejiang Province Key Laboratory of Quantum Technology $\&$ Device and Department of Physics, Zhejiang University, Hangzhou 310027, P.R. China.}
\author{You-Quan Li}\email[email: ]{yqli@zju.edu.cn}
\affiliation{\noindent
Zhejiang Province Key Laboratory of Quantum Technology $\&$ Device and Department of Physics, Zhejiang University, Hangzhou 310027, P.R. China.}
\affiliation{\noindent
Collaborative Innovation Center of Advanced Microstructure, Nanjing University, Nanjing 210008, R.R. China.}


\received{\today}

\begin{abstract}
The idea that a search efficiency can be increased with the help of a number of autonomous agents is often relevant in many situations, which is known among biologists and roboticists as a stigmergy.
This is due to the fact that, in any probability-based search problem,
adding information provides values for conditional probabilities.
We report new findings of a speed-up and suppression effects
occurring in the quantum search problem
through the study of quantum walk on a graph
with floating vertices.
This effect is a completely counterintuitive phenomenon in comparison to the classical counterpart£¬
and may facilitate new insight in the future information search mechanisms
that were never been perceived in classical picture.
In order to understand the first passage probability,
we also propose a method via ancillary model to bridge the measurement of the time dependence of the total probability of the complimentary part and the first passage probability of the original model.
This is expected to provide new ideas for quantum simulation by means of qubit chips.
\end{abstract}

\maketitle

Searching and foraging has long been crucial topics in the ecology and microbiology~\cite{Shlesinger2006,Benichou2006},
it becomes enormously important in physics as well as in information science nowadays~\cite{Foulger2014}.
The idea that a search efficiency can be increased
with the help of a number of autonomous agents
was often relevant in many situations.
Swarm communication is widely adopted among animals~\cite{Bell1990} and insects~\cite{Dethier},
and it is known among biologists~\cite{Ariel2015} and roboticists~\cite{Fujisawa} as a stigmergy.
For example, in the search strategies of some predatory animals, which mix essentials
of random-walk model with concentration on seasonal hot spots to find preys~\cite{Benichou2014}.
Similarly, in biological cells, peptide binding to transmembrane receptors relies on hydrophobic
attraction superimposed on random Brownian motion~\cite{Kennedy}.
L\'evy  walks are characterized by trajectories that have straight stretches for extended lengths whose variance is infinite~\cite{Klafter1996}.
Comparing to classical random walks, the L\'evy  walks were shown
to being advantageous in optimality of search~\cite{Shlesinger1986,Viswanathan}.
Group L\'evy foraging with an artificial pheromone communication between robots
was studied~\cite{Fujisawa}
where the experimental results showed that
if an interaction or exchange of information between the searchers is allowed
the averaged search time can be decreased substantially.
All those situations will continue to provide various questions for theorists~\cite{Zaburdaev}.

Recently,
modelling protein folding as a quantum walk on definite graphs reveals a fast protein-folding time~\cite{LuLi2019}.
This motivate us raise a general question
whether quantum walk strategy can provide us any new insight on the searching efficiency~\cite{Tang2018}.
Here we show our new findings that there are speed-up and suppression effects
occurring in the quantum search problem that does not occur in its classical counterpart.
We introduce a bilayer graph in which a layer of switchable vertices is ¡®floating¡¯
over a main chain of vertices.
We study the cases of the main chain only and compare successively with the case of one- and two-vertex on side chain connecting to the main chain.
The mean first passage time starting from one terminal
to the other terminal are calculated and the effect caused by connecting one or two vertices
from the side-chain are investigated.
As the search efficiency can be quantitatively measured by the mean first passage time~\cite{Friedman2017,Muga2000},
we investigate various situations of our model
in the framework of the classical random walk and quantum walk, respectively.
We also introduce reduced density matrix and calculate the von Neumann entropy
to get certain knowledge about the counterintuitive effect.
In order to understand the first passage probability,
we also propose a method via ancillary model to bridge the measurement of the time dependence of the total probability of the complimentary part and the first passage probability of the original model.
Quantum search's feature is due to its parallel calculation strategy
and quantum coherence nature.
It is expected to provide new ideas for quantum simulation by means of qubit chips.

\begin{figure}[h]
\centering
\vspace{-1mm}
\includegraphics[width=0.7\linewidth]{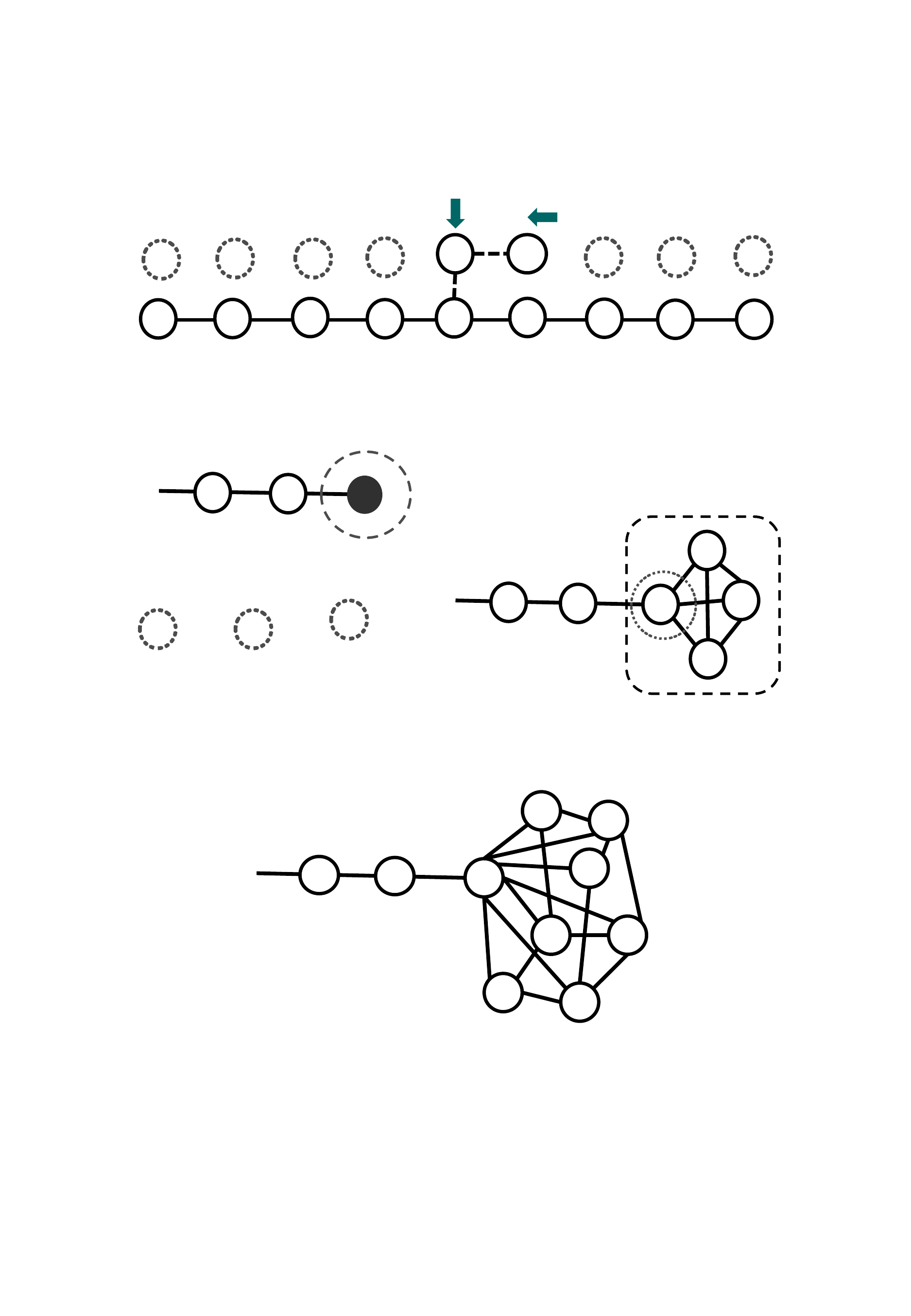}
\vspace{-1mm}
\caption{{\bf Graphical illustration of the model Hamiltonian.}
Here the dash-line circle denotes isolated vertex in the float layer
that can be pushed to connect with a vertex in the base layer.
If another neighbor in the float layer is sheared to connect with,
a side chain of two vertices is tuned on then.}
\label{fig:model}
\end{figure}

\paragraph{Model system and the mean first passage time:}
We consider a two layer system that consists of a back layer (base layer) of a chain of $N$ vertices
and a front layer (float layer) of several initially isolated vertices
which are switchable to connect with the contact-vertex in base layer.
In the float layer, another neighbor vertex can also be switched on or off from the aforementioned connected vertex in the float
layer (Fig.~\ref{fig:model}).
\noindent
Thus, we investigate a model of
a main chain of $N$ vertices along with a side chain of $S$ vertices mounted at the center.
In the model, the number of the side chain can be changed by tuning up one more vertex,
that means the $S$ can be changed  from odd number to even number, or vise versa.
We explore how does the side chain affect
the mean first passage time that costs from initial site, saying vertex-$1$,
to target site, saying vertex-$N$.
The mean first passage time is defined
as an integrand, namely,
\begin{equation}\label{eq:mean}
\tau=\int_0^{\tau_0} t F_{1,N}(t)\mathrm{d}t
 \Bigl/ \int_0^{\tau_0} F_{1,N}(t)\mathrm{d}t,
\end{equation}
where $\tau_0=\infty$ in the classical random walk while it ought to be determined by $F_{1,N}(\tau_0)=0$ in quantum-walk problem~\cite{LuLi2019}.
Here $F_{1,N}(t)$ is determined by the convolution relation~\cite{FPT1969,FPT2001,FPT2004,FPT2007,FPT2016},
\begin{equation}
P_{1,N}(t) =\int^t_0 F_{1,N}(t') P_{N,N}(t-t')\mathrm{d}t',
\label{eq:convolution}
\end{equation}
where $P_{a,b}(t)$ is the time-dependent probability distribution at vertex-$b$
evolving from a initial distribution located merely  at vertex-$a$, i.e.,
$P^{}_{a,b}(0) =\delta_{a b}$.

\paragraph{The classical random walk:}
Random walks on a graph (Fig.~\ref{fig:model})
is described by the probability distribution over the vertices, namely,
$p^{}_{a}(t)$ with $a=1, 2, \cdots, N, N+1, N+2, \cdots N+S$
obeying  the following master equation,
\begin{equation}\label{eq:master}
\frac{\mathrm{d}}{\mathrm{d}t} p^{}_a (t) =\sum_b K^{}_{a b}p^{}_b(t).
\end{equation}
Here $K_{a b}=J_{a b}/\mathrm{deg}(b)-\delta_{a b}$
with $J_{a b}$ being the adjacency matrix of the graph and
$\mathrm{deg}(b)=\sum_c J_{c b}$ the degree of vertex-b.
Let us first observe the classical random walk on the graph (Fig.~\ref{fig:model})
for $S=0,$ $1$ and $2$  one by one.
For each case, we solve the master equation (\ref{eq:master}), respectively,
under the initial condition $p^{}_1(0)=1$ and another initial condition $p^{}_N(0)=1$.
Those solutions provide us
$P_{1,a}(t)$ and $P_{N,a}$ that determine the so-called first passage probabilities~\cite{FPT1969,FPT2001,FPT2004,FPT2007,FPT2016} $F_{1,N}(t)$
through the convolution relation (\ref{eq:convolution}).
Strictly speaking, the $F_{a,b}(t)$ measures a probability per unit time.
After solving the time dependence $F_{1,N}(t)$, we can evaluate the mean first passage time as an integrand defined by Eq.~(\ref{eq:mean}).
\begin{figure*}[th]
\centering
\vspace{0mm}
\includegraphics[width=0.90\linewidth]{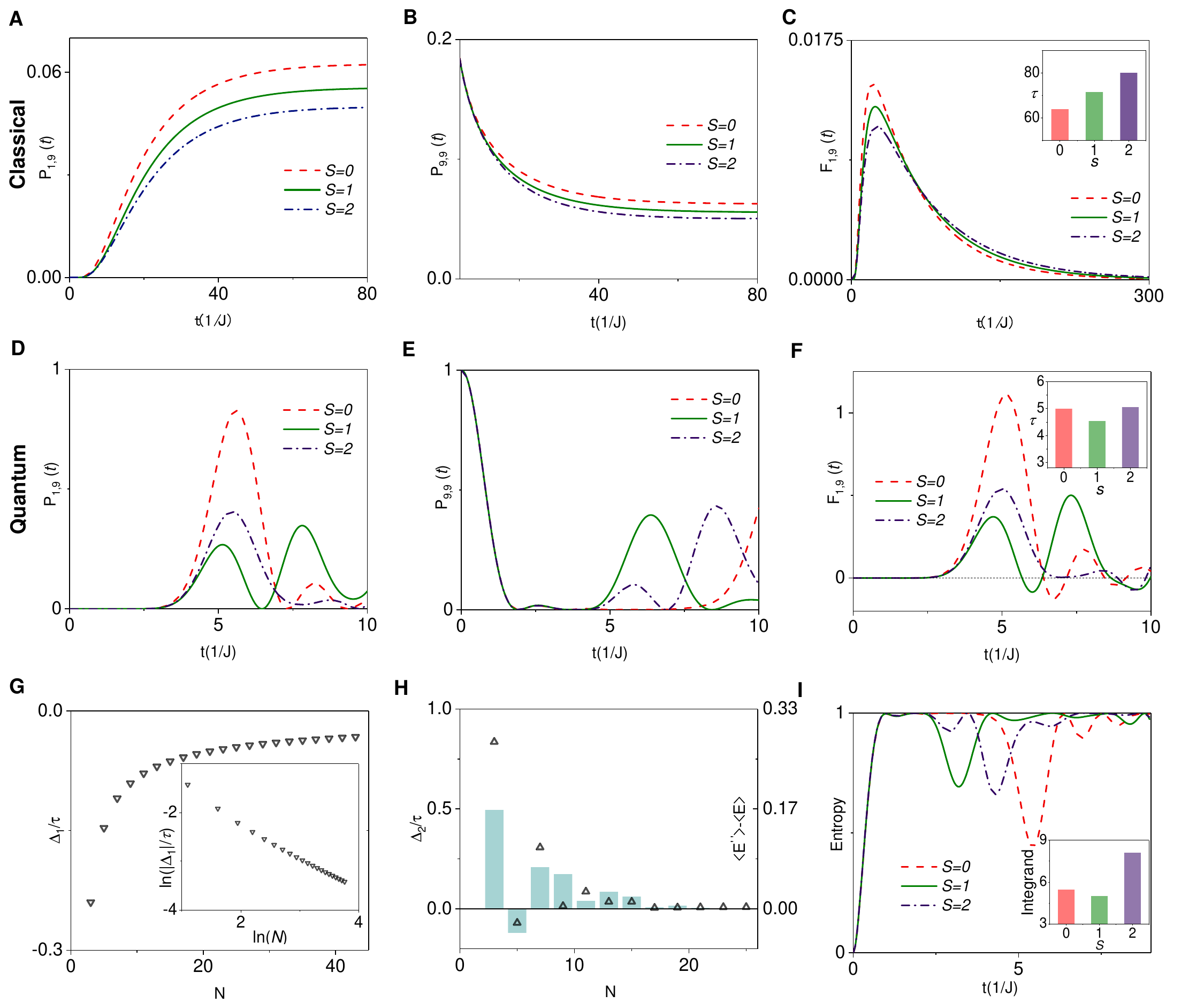}
\vspace{-1mm}
\caption{\textbf{Classical and quantum solutions, their features of the mean first passage time:}
The time dependence of
(\textbf{A}) $P_{1,9} (t)$,
(\textbf{B}) $P_{9,9}(t) $ and
(\textbf{C}) $F_{1,9} (t)$ solved from classical random walk;
the time dependence of
(\textbf{D}) $P_{1,9} (t)$,
(\textbf{E}) $P_{9,9}(t) $ and
(\textbf{F}) $F_{1,9} (t)$ solved from quantum walk for $N=9$.
(\textbf{G}) The quantum speed-up ratio $\Delta_1/\tau$ versus the number of sites $N$,
the inserted panel is the logarithm scaled plot indicating a power low behavior.
(\textbf{H}) The relative suppression ratio $\Delta_2/\tau$ versus the number of sites $N$, presented by
small triangular,
together with the magnitude difference between the averaged von Neumann entropies for $S=2$ and $S=0$
presented by histogram which is scaled by the right vertical axis.
(\textbf{I}) The time dependence of von Neumann entropies for $S=0, 1, 2$  when $N=9$.
Their integrands are shown in the inserted panel.
}
\label{fig:time}
\end{figure*}

We numerically solve the random walk on systems with different number of vertices on the main chain,
saying
$N=3$, $N=5$, $\cdots$, till $N=43$.
The solution of $N=9$ is plotted in Fig.~S1
and the relevant quantities for solving the first passage probabilities are plotted
in Fig.~\ref{fig:time}.
We calculate the mean first passage time
from vertex-$1$ to vertex-$N$,
and denote the obtained value for $S=0$ by $\tau_\mathrm{c}$, that for $S=1$ by $\tau'_\mathrm{c}$ and that for $S=2$ by $\tau''_\mathrm{c}$.
Here the subscript ``c'' refers the magnitude from classical random walk.
Our results exhibit that
$\tau'_\mathrm{c}$ is always larger that $\tau_\mathrm{c}$
and $\tau''_\mathrm{c}$ is always larger that $\tau'_\mathrm{c}$
(Fig.~\ref{fig:time}C and Table S1).
This manifests that an additional vertex connected from the float layer always  retards the mean first passage time~\cite{FPT2001}
of classical random walk in some sense, which just fits with the daily conventional intuition.

\paragraph{Quantum severalty:}
The quantum walk~\cite{Aharonov1993,Farhi1998,WangJB2014} on a graph is described by the time-dependent
wavefunction
 ${\mid\!\Psi(t)\rangle}=\sum_{a=1}^{N+S}\psi_a(t)\mid\!a\,\rangle$
that obeys the Schr\"odinger equation,
\begin{equation}
i\hbar\frac{\mathrm{d}}{\mathrm{d}t}\mid\!\Psi(t)\rangle
 =\hat{H}\mid\!\Psi(t)\rangle.
\label{eq:Sch-eq}
\end{equation}
Here the Hamiltonian $\hat{H}=\sum_{a b}J_{a b}\mid\!a\,\rangle \langle\,b\!\mid$
is defined by the adjacency matrix $J_{ab}$ of the graph.
After solving Eq.~(\ref{eq:Sch-eq})
under the initial condition $\mid\!\Psi(0)\rangle=\mid\!1\,\rangle$ and another initial condition $\mid\!\Psi(0)\rangle =\mid\!N\,\rangle$,
respectively,
we get
$P_{1,N}(t)=|\psi^{\small (1)}_N(t)|^2$ and $P_{N,N}(t)=|\psi^{\small (N)}_N(t)|^2$
(the superscript is introduced to distinguish solutions from different initial conditions),
then obtain $F_{1, N}(t)$ from the convolution relation (\ref{eq:convolution}).
We solve the quantum mechanical problem for the cases $S=0$, $1$ and $2$ one by one
and furthermore evaluate the mean first passage time in terms of the solved $F_{1,N}(t)$.
The time dependence of $P_{1,N}(t)$, $P_{N,N}(t)$  and $F_{1,N}(t)$ in quantum case for
$N=9$ are plotted in Fig.~\ref{fig:time}D-\ref{fig:time}F.

The calculated mean first passage time is given in the inserted panel of Fig.~\ref{fig:time}F.
We can see $\Delta_1=\tau' -\tau <0$,
which implies that
the mean first passage time for $S=1$ is shorter than that for $S=0$.
Unlike the classical case where an additional vertex at side chain will increase the mean first passage time (retard the searching rapidity),
quantum mechanically, it will speed up the searching rapidity (decrease the mean first passage time).
Our result from the quantum walk reveals a novel effect that is a completely counterintuitive phenomenon.
In comparison to its classical counterpart, we suggest to call it ``quantum severalty''.
This is true not only for $N=9$, but also for
other number of vertices.
We also calculate $N=3, 5,\cdots,~\textrm{till}~ 43$ (Table S1)
and plot the speed-up ratio $\Delta_1/\tau$ as a function of $N$, the number of the vertices
on the main chain (Fig.~\ref{fig:time}G).
The fitted curve fulfills a power low behavior (see the inserted panel of Fig.~\ref{fig:time}G),
namely,
$$
\frac{\Delta_1}{\tau} \approx -0.4574 N^{-0.714}.
$$

If connecting one more vertex to the already connected one on the side chain,
we find that $\Delta_2=\tau'' -\tau$ turns to be a tiny magnitude.
This means tuning on a second vertex on the side chain,
the original speed up effect will be suppressed down at once.
The suppression magnitude $\Delta'_2=\tau''-\tau'=\Delta_2-\Delta_1 >0$.
The relative suppression ratio
$\Delta_2/\tau$ versus the site number $N$ is shown in Fig.~\ref{fig:time}H.
Thus, if one float vertex is pushed to connect with the base layer, we attain a significant  speed up effect;
furthermore, if a second nearby float vertex  is sheered to
connect with the connected float vertex, it causes a suppression effect.
It is also worthwhile to know what happens if the side chain is not mounted at the center of the main chain.
Denoting the central position as $c$,
we discuss the case when the side chain is on the position $c\pm 1$, $c\pm 2$ etc..
Our calculation results exhibit that the speed-up and suppress effect still exists if the
the side chain is mounted nearby the center, and the magnitude changes are just affected slightly (Table S2).

To help an understanding about the aforementioned quantum severalty,
we consider a reduced 2 by 2 density matrix $\tilde{\rho}$ in the spirit of Ref.~\cite{MaoLuLi}
that may maintain certain information of quantum coherence.
Because the graph characterising our model system is actually a two-color graph
(i.e., minimally, two colors are needed to dye every vertices without the occurrence of neighbor vertices sharing the same color),
we are able to define the 2 by 2 density matrix $\tilde{\rho}$ in terms of
$\tilde{\rho}^{}_{1 1}=\sum_{a_1}\psi_{a_1}^{}\psi_{a_1}^*$,
$\tilde{\rho}^{}_{2 2}=\sum_{a_2}\psi_{a_2}^{}\psi_{a_2}^*$,
$\tilde{\rho}^{}_{1 2}=\sum_{a_1 a_2}\psi_{a_1}^{}\psi_{a_2}^*/\sqrt{n_1 n_2}$
and
$\tilde{\rho}^{}_{2 1} = \tilde{\rho}_{1 2}^*$
where
$n_1$ and $n_2$ stand for, respectively, the total numbers of the vertices in the same color
and the square-root-denominator factor in the summation guarantees
the reduced density matrix such defined
is of non-negatively definite.
Then the von Neumann entropy~\cite{Nielsen2001} is given by
$E(\tilde{\rho})\,=\,-\mathrm{tr} (\tilde{\rho} \,{\rm {\log_2}}\tilde{\rho})$
whose time dependence for $N=9$ is plotted in Fig.~\ref{fig:time}I.
We denote the von Neumann entropies for $S=0, 1, 2$ respectively by
$E$, $E'$ and $E''$,
of which the integrand over the whole period $\tau_0$ is indicated by histograms in the inserted panel.
One can also make an average of the von Neumann entropy,
$\langle E \rangle=\int^{\tau_0}_0 E(t)\mathrm{d}t/\tau_0$,
clearly, the difference
$\langle E''\rangle - \langle E\rangle$ have something to do with the
$\Delta_2/\tau$-$N$ dependence (Fig.~\ref{fig:time}H).
All these features are true also for $N=3, 5, \cdots $ (Fig.~S2).
\begin{figure}[t]
\centering
\includegraphics[width=0.38\textwidth=0.95]{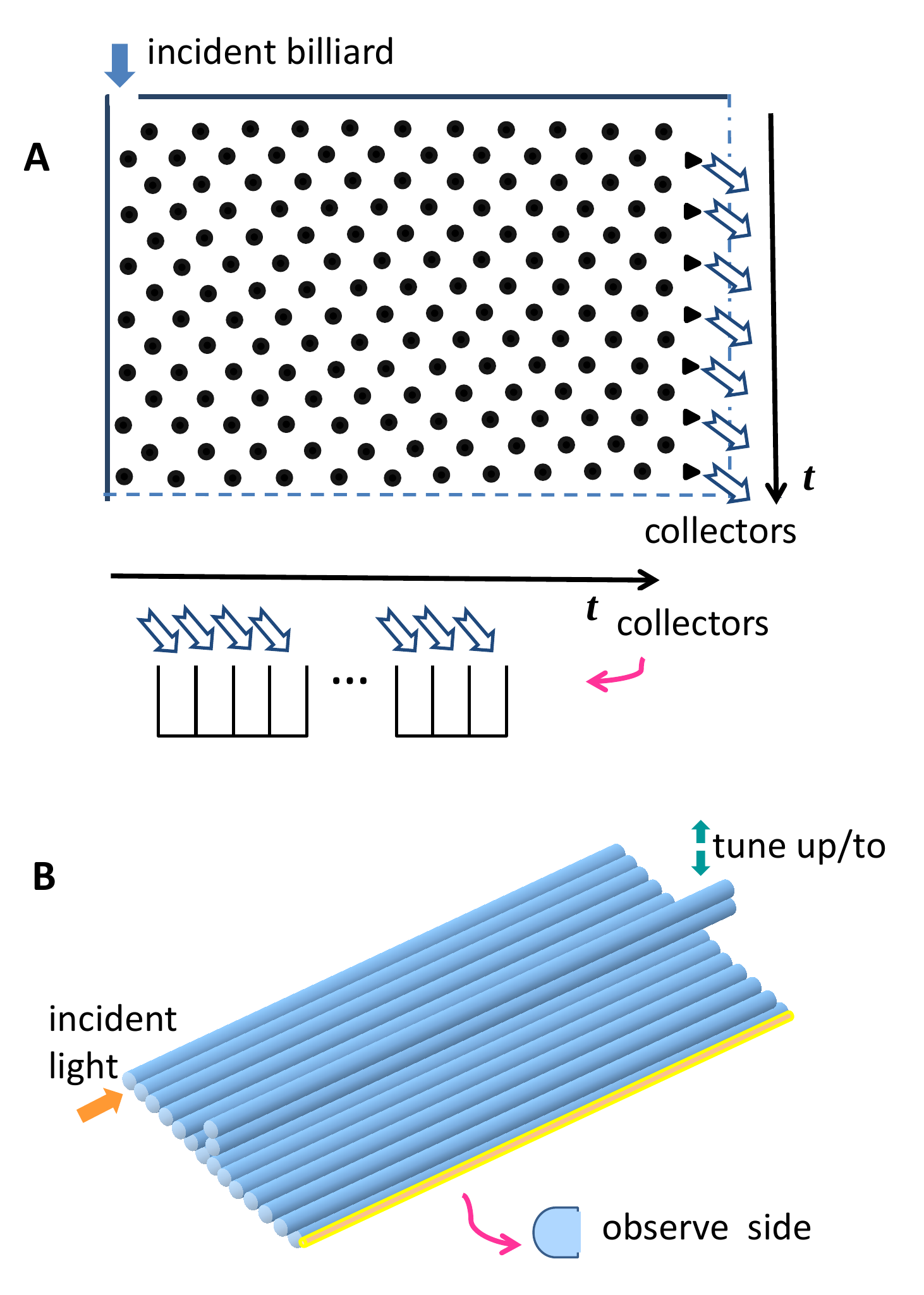}
  \vspace{-2mm}
\caption{\textbf{Proposals for experiment set-up}:
(\textbf{A}) Demonstration experiment set-up for random walk.
This set-up is mounted vertically so that the gravitational force naturally provide us the time order direction.
The collectors on the right-side are labelled by this time order.
(\textbf{B}) Experiment scheme for quantum walk via optical fibre.
Incident beam of light is applied from one side and the observation is made
on the other side of which the lateral fibre is cut to form a unreflecting surface.
The speed up effect, and furthermore a suppress effect will be observed by tuning up or tuning to the additional fibre.
}
\label{fig:setup}
\end{figure}
\begin{figure}[h!]
\centering
\vspace{2mm}
\includegraphics[width=0.9\linewidth=0.95]{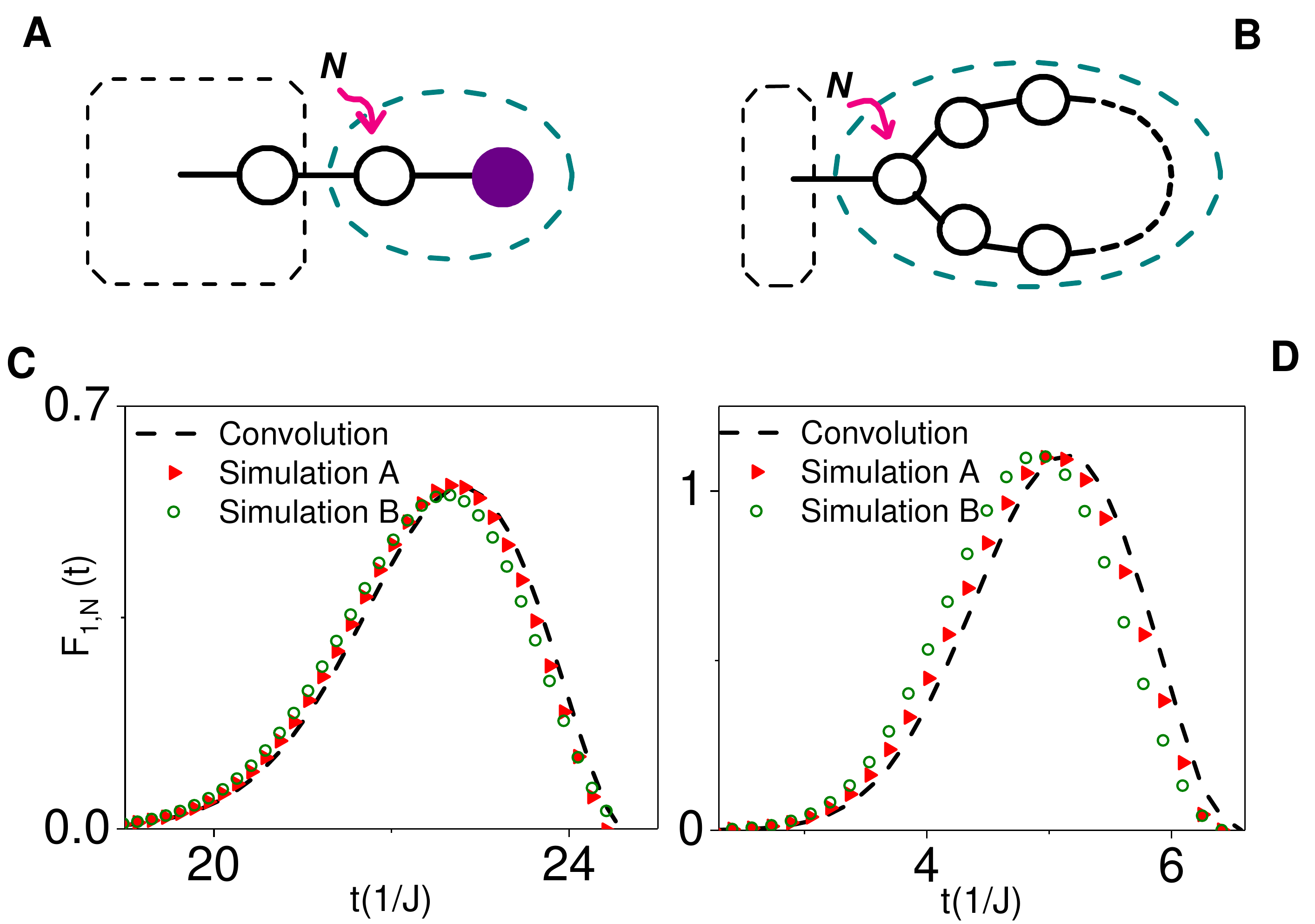}
\vspace{-2mm}
\caption{\textbf{Quantum simulation scheme via qubit chip}:
(\textbf{A}) Ancillary model with a sticky tail, which is obtained by connecting to the $N$-th vertex with an additional sticky vertex;
(\textbf{B}) Ancillary model dressed with a ring, which is a prolongation of the $N$-th vertex to be a ring of several more vertices;
(\textbf{C}) The first passage probability obtained, respectively, by means of the above proposals for $N=43$, and
(\textbf{D}) that for $N=9$. The datas obtained via convolution relation are also plotted to compare.}
\label{fig:simulation}
\end{figure}

\paragraph{Implications of the first passage probability:}
Classically, the first passage probability $F(t)$ can be obtained directly via ensemble simulation.
A numerical simulation of one million ensembles gives rise to a first-passage probability that fit well
with the convolution results (Fig.~S3).
We know that the one dimensional random walk can be demonstrated by  billiards scattered by periodically placed nails (Fig.~\ref{fig:setup}A).
In order to understand the physics implication of the first passage probability $F_{1,N}(t)$,
we make some changes on the set-up frequently adopted in science and technology museum.
Pouring in billiards from the hole close to the left edge,
keeping the right-side boundary open and placing an array of collectors to receive,
successively,
the billiard scattered toward outside,
we will have a $F_{1,N}(t)$ after sufficient billiards were poured in.
The magnitudes of $F_{1,N}(t)$ is valued by accounting the billiards
received by those collectors labelled by the $t$.
Thus, our simulation (Fig.~S3)
on classical random walk can be realized by such a demonstration experiment (Fig.~\ref{fig:setup}A).

The aforementioned classical set-up motivated us, at once, to propose a quantum simulation experimental scheme.
One can make a set-up (Fig.~\ref{fig:setup}B) in terms of optical fibre~\cite{Perets}
by cutting a lateral fibre to form a reflectionless surface,
then the observed intensity along the fibre will be the $F_{1,N}(t)$.
One can also observe the new phenomena we found previously~\cite{dark},
i.e., the speed up and suppress effects caused by tuning to or tuning up side-chain optical fibre.

\paragraph{Quantum simulation by qubit chips:}
We propose possible experiment design (Fig.~\ref{fig:simulation})
to carry out quantum simulations~\cite{quSimulation2014} via qubit chips.
To measure the first passage probability $F_{1,N}(t)$ at vertex-$N$,
we can either connect an additional sticky vertex to that vertex (Fig.~\ref{fig:simulation}A)
or make a prolongation of that vertex to be a ring of several more vertices (Fig.~\ref{fig:simulation}B).
For the former proposal (Fig.~\ref{fig:simulation}A),
as we know, in the presence of sticky, we need to solve the density matrix from
the Lindblad equation~\cite{Lindblad}
\begin{equation}
\frac{\mathrm{d}}{\mathrm{d} t}\tilde{\rho}
  = \frac{1}{i\hbar} [\tilde{H},\tilde{\rho} ]
   +\frac{\lambda}{2}\bigl(
 2 L\tilde{\rho}\,L^{\dagger}
    -\tilde{\rho}\,L^{\dagger}L
    -L^{\dagger}L \tilde{\rho}
    \bigr),
\label{eq:Lindblad}
\end{equation}
where
$\tilde{H}=\hat{H} + V\mid\!{\small N+1\,}\rangle\langle\,{\small N+1}\!\mid$ with $V$ being the negative potential on the sticky vertex,
and
$L=\mid\!N\,\rangle\langle\,N+1\!\mid$ the Lindblad operator.
Once the density matrix $\tilde{\rho}_{a b}(t)$ ($a, b=1, 2, \cdots, N, N+1$) is solved
for the ancillary model with a sticky tail (Fig.~\ref{fig:simulation}A),
we are able to obtain the first passage probability:
\begin{equation}\label{eq:theF}
F_{1,N}(t)=-\frac{1}{A}\frac{\mathrm{d}}{\mathrm{d}t}\sigma (t),
\end{equation}
where
$\sigma(t)=\sum_{a=1}^{N-1}\tilde{\rho}_{aa}(t)$
and $A$ is a normalization constant so that
$\int_{0}^{\tau_0}F_{1,N}(t)\mathrm{d}t=1$,
i.e.,
$A=\sigma (\tau_0)$.
If the realization of the sticky vertex in some experimental system is uneasy,
an alternative proposal is to prolong the vertex $N$ to be a ring of several qubits (Fig.~\ref{fig:simulation}B),
we call it the ancillary model dressed with a vertices ring.
With this ancillary model, one is still able to simulate the $F_{1,N}(t)$ by the same formula (\ref{eq:theF}), but,
with the $\sigma (t)$ given by
$\sigma(t)=\sum_{a=1}^{N-1}|\tilde{\psi}^{(1)}_a(t)|^2(t)$.
Our quantum simulation results for $N=43$
together with the convolution result are plotted in
Fig.~\ref{fig:simulation}C,
where the parameter choices are
$\lambda=4.6$, $V=-2.3J$
and the ring of $44$ vertices.
Fig.~\ref{fig:simulation}D shows that for $N=9$
with parameter choices $\lambda=5$, $V=-2.5J$ and the ring of $10$ vertices.
Therefore, there is a bridge between the probability measurement of the ancillary model
and the first passage probability of the original model, which will shed new light on the
research area of quantum simulation.

This work is supported by National Key R \& D Program of China, Grant No. 2017YFA0304304
and NSFC, Grant No. 11935012.

\bibliography{50}

\begin{thebibliography}{50}
\bibitem{dark}
Even more, if the optical fibres are sufficient long, there may exists a `dark period'
which we mean the time interval that the solved first passage probability
becomes negative.
\end{thebibliography}

\appendix

\end{document}